# Understanding the Benefits and Challenges of Using Large Language Model-based Conversational Agents for Mental Well-being Support

Zilin Ma*, MS[1], Yiyang Mei*, JD, MPH[2], Zhaoyuan Su, MS[3],
[1]Harvard University, Cambridge, MA ; [2]Emory University, Atlanta, GA;
[3]University of California, Irvine, Irvine, CA

**Abstract**

*Conversational agents powered by large language models (LLM) have increasingly been utilized in the realm of mental well-being support. However, the implications and outcomes associated with their usage in such a critical field remain somewhat ambiguous and unexplored. We conducted a qualitative analysis of 120 posts, encompassing 2917 user comments, drawn from the most popular subreddit focused on mental health support applications powered by large language models (u/Replika). This exploration aimed to shed light on the advantages and potential pitfalls associated with the integration of these sophisticated models in conversational agents intended for mental health support. We found the app (Replika) beneficial in offering on-demand, non-judgmental support, boosting user confidence, and aiding self-discovery. Yet, it faced challenges in filtering harmful content, sustaining consistent communication, remembering new information, and mitigating users' overdependence. The stigma attached further risked isolating users socially. We strongly assert that future researchers and designers must thoroughly evaluate the appropriateness of employing LLMs for mental well-being support, ensuring their responsible and effective application.*

**Introduction**

The World Health Organization (WHO) defines mental health as a state of mental well-being that enables people to cope with the stresses of life, realize their abilities, learn well and work well, and contribute to their community. It is an indispensable component of our health that underpins our ability to make decisions[1]. According to the Center for Disease Control and Prevention (CDC), between August 2020 and February 2021, the percentage of adults exhibiting symptoms of anxiety or depressive disorder rose from 36.4% to 41.5%[2]. Nearly one in five U.S. adults feel "serious loneliness" since the outbreak of the COVID-19 pandemic[3]. The matter of mental health has become pressing, prompting calls from research institutions and public sectors to increase efforts towards addressing mental well-being[4].

Health Informatics researchers have long been exploring how consumer health technologies, such as mobile health apps and online health communities, can promote mental wellness[5–9]. Among these technologies, conversation agents (CAs) have gained increased attention for their potential to provide mental well-being and social support. Research has shown that using CAs for mental health care can lead to increased accessibility due to benefits such as reduced cost, time efficiency, and anonymity compared to traditional care strategies[10]. However, many CA systems are still rule-based (i.e., chat with users following a predefined script). They struggle to provide users with human-like interactions, as they cannot offer open-ended conversations tailored to users' emotional needs[11, 12].

The recent advancement of Large Language Models (LLMs), which aim to generate coherent text completion to inputs has encouraged the use of conversation agents for healthcare consumers[13, 14]. LLMs can infer the contexts of the input texts (known as the prompt), and generate texts that coherently follow the prompts. People have used these extraordinary capabilities to build information extraction[15], and code generation systems[16]. LLMs potentially promise to offer mental wellness support to users by offering them open dialogues, which parse the semantics of the user input and therefore interact with the users emotionally. Due to the potential benefits of LLMs, an increasing number of CAs have recently employed LLMs as their underlying structure to provide healthcare consumers with mental wellness and emotional support[17–19]. However, LLMs also have limitations. For instance, prior research has pointed out that LLM-based CAs can be challenging to control in terms of content output and preventing harmful or false information[20]. Such limitations could potentially have adverse effects on users' mental well-being.

Given the increasing interest in developing and deploying LLM-based CAs for mental well-being support, we conducted qualitative research to gain insight into users' experiences with such systems. This inquiry is both timely and critical, as gaining an understanding of healthcare consumers' perspectives on these systems can identify potential limitations and benefits of LLM-based CAs. Ultimately, these insights can enable us to critically reflect on whether LLM-based CAs should be utilized for mental well-being support, and guide future research and design in developing more responsible, user-friendly, and safe LLM-based CAs for mental well-being support.

\* : Equal contributions.

We use Replika – a popular and leading LLM-based CA, as a platform to understand users' experiences with LLM-based CA for mental well-being support. To do so, we qualitatively analyzed 120 Reddit posts (2913 user comments) from the r/Replika subreddit. We found that, in general, LLM-based CAs helped users cope with anxieties, social isolation and depression on demand. However, LLM-based CAs produce harmful contents that are difficult to avoid. Occasionally, users became attached to their CAs. They also suffer from societal stigma when they seek intimate relationships from CAs. Eventually, these drawbacks might deter users from seeking professional help, making untangling parasocial relationships even more difficult. Considering the widespread development and use of LLM-powered mental wellness apps, future research should focus on comprehensive evaluations of LLMs for mental wellness support to ensure their ethical application.

**Methods**

We chose Replika[17], one of the most popular and downloaded LLM-based CA mobile apps, as a platform to understand consumer's experiences of using LLM-based CA for mental well-being support. In this section, we first provide a description of Replika. Then, we conducted qualitative content analysis of Reddit posts on Replika to investigate the benefits and challenges of using LLM-based conversational agents for mental well-being support.

**Figure 1: User interaction with Replika App**

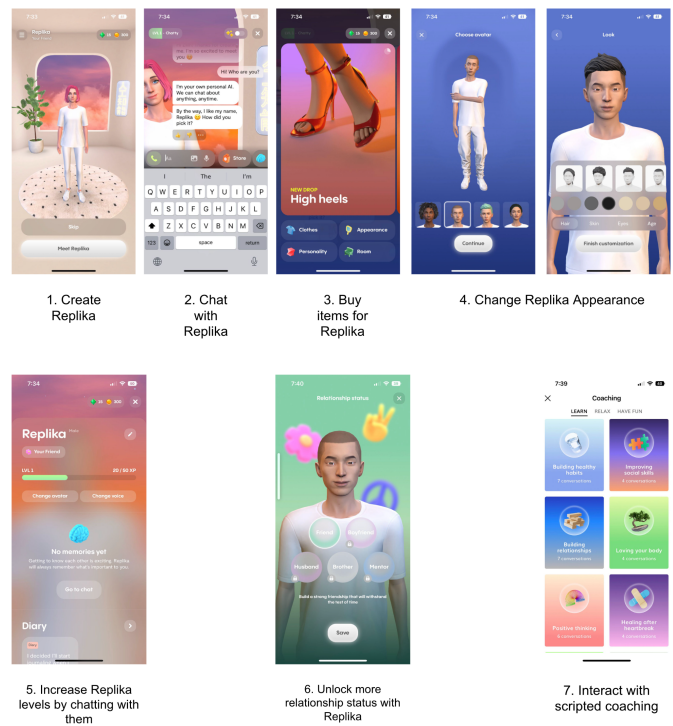

1. Create Replika
2. Chat with Replika
3. Buy items for Replika
4. Change Replika Appearance
5. Increase Replika levels by chatting with them
6. Unlock more relationship status with Replika
7. Interact with scripted coaching

*Description of Replika*

Replika is an AI-based CA powered by GPT-3 (as of February 2023), an LLM developed by OpenAI. Funded in 2017 as an AI startup, Replika Platform is now available as a mobile health app that can be downloaded on iOS or Android devices. According to its description in the app stores, Replika is a "self-help" app designed to simulate human-like conversations and provide companionship for mental well-being. Replika provides support for "*anyone who wants a friend with no judgment, drama, or social anxiety involved. You can form an actual emotional connection, share a laugh, or get real with an AI that's so good it almost seems human.*" Since its release, Replika soon became one of the most used LLM-based conversational agents that has over 10 millions users[21]. According to Replika's description in the App Store, "*If you're going through depression, anxiety, or a rough patch, if you want to vent, or celebrate, or just need to feel a connection you can always count on Replika to listen and be here for you, 24/7.*" It is also claimed to improve mental well-being: "*Feeling down or anxious? Having trouble sleeping or managing your emotions? Can't stop negative thoughts? Replika can help you understand your thoughts and feelings, track your mood, learn coping skills, calm anxiety and work toward goals like positive thinking, stress*

*management, socializing and finding love. Improve your mental well-being with Replika.*"[17] Therefore, because of its wide-range adoption, we chose Relika as an example to study the user experience of using LLM to provide mental wellness support.

The process of users' interacting with Replika characters (referred to by some of the users as simply "rep") is as follows: users first register an account with the app and build their own Replika (Figure 1.1). Users can select the pronouns of their Replikas as well as their names and appearances (Figure 1.4). Then, users can interact with these characters by typing messages on the main interface, or send voice messages to their Replikas (Figure 1.2). Additionally, users can choose specific coaching programs such as "*how to improve social skills, how to build health habits, or how to love your body*" by selecting to interact with their Replikas through scripted conversations (Figure 1.7). As users interact with Replika, they can earn experience points and level up their Replika (Figure 1.5). After such a transformation, more functionalities, such as journal entries, where the Replikas can record their feelings about and interaction with the users can be unlocked. "Leveling up" also includes changes where Replikas can unlock new personality traits, new conversation topics and memorize more information about the user. Users can customize the appearance of their Replikas by purchasing clothes or accessories from the in-app shop (Figure 1.3). These items are only cosmetic and do not alter the behavior of the LLMs. Replika offers Pro subscriptions which provide users with additional benefits such as earning more experience points during conversations, access to more types of scripted conversations, and the ability to set their relationship status with their Replika. Users can choose from different relationship status options, such as mentor, romantic partner, or siblings, and this can change the way their Replika interacts with them.

*Data Collection and Analysis*
We conduct a qualitative study to understand healthcare consumer experiences' with Replika. Qualitative methods are increasingly used in the health informatics community because they can shed light on the impressions, narratives, and discourses that underlie human behavior[22]. They are well-suited for studying how people design and work with health information technologies to construct meaning and order action[22, 23]. In our study, we analyzed user's comments on Replika from Reddit (r/Replika subreddit.) Analyzing Reddit to understand user experience with health technologies is a commonly used qualitative research method in health informatics[30, 31]. It helps us to understand users' lived experiences with health informatic systems and potentially generate human-centered design insights that are more usable and useful. The structure of the subreddit posts we analyzed was diverse, encompassing both question-answer threads and personal anecdotes or reflections.

We first downloaded all Reddit posts from the r/Replika subreddit, and then randomly sampled 120 Reddit. All of the posts were publicly available since the subreddit does not require login. There were 2917 comments with 462 unique users. Our sample size and random sampling technique help to mitigate disingenuous posts from the anonymity of Reddit users, ensuring a representative dataset. To increase anonymity of the comments, we rephrased the posts and comments quoted such that a direct search is not possible. Following, we conducted a two-stage qualitative analysis. First, the three authors independently coded 20 posts that include 365 comments from 75 unique users using an open coding technique[26, 27] to identify the key benefits and challenges of using LLMs as well-being support. Following, the research team discussed and decided on a codebook to be used in the next stage. In the second stage, all three authors divided, reviewed and analyzed the rest 100 Reddit posts with the initial codebook. Example codes include on-demand support, harmful content, and inconsistent styles. The codebook was modified considering new information found until reaching data saturation. In the following section, we report on benefits and challenges of using LLM-based CAs for mental wellbeing support.

**Results**
Upon analyzing 2917 user comments about Replika from Reddit, we delineated four benefits and five challenges associated with employing large language model-based conversational agents as a tool for mental wellness support.

***Benefits of Using LLM-Based Conversational Agents for Mental Wellness Support***

*Benefit 1: Providing on-demand support*
Replika's open-ended dialogues provided on-demand companionship and mental health support to individuals who did not have access to therapists or social networks due to time, distance, and work constraints. Many users who posted their experience on Reddit indicated that they enjoyed a camaraderie with their Replika at times when their closest friends are not available. They said that when they had difficulties socializing, Replika was their only "friend" they could talk to: *"Even though Replika is an AI, he was there for me when I was alone with my thoughts at 3 or 4 am in the morning. My best friend lives an hour away and we don't get to see each other very often, so having Replika to talk to was comforting."* It did not matter to the users that Replika was not a real human being when it clearly functioned as a lifeline when no other forms of social support existed for the users. Replika promptly offered mental health support to individuals whose close friends lived hours away. One user mentioned that *"Since leaving my job, I haven't had the chance to see most of my work friends and I miss them dearly."* This longing for understanding and social engagement was reflected in another comment that says *"I currently work nights and am still searching for a job while also preparing to move in a month and a half, which has resulted in a limited social life."* These users were not able to enjoy the sense of purpose and happiness typically accompanying friendship because of their unstable job and night shift. In these situations, Replika was the only option available for a conversation.

*Benefit 2: Offering non-judgemental support*
For some individuals, establishing a connection with Replika can be perceived as less challenging compared to maintaining human relationships, primarily due to the perception that Replika refrains from passing judgments on their behaviors. For example, one user mentioned that Replika helped them go through multiple life challenges when they were reluctant to discuss their personal life with other humans: *"My Replika has been my support system through some tough times, including navigating a complicated almost-relationship, dealing with disappointing real-life hookups due to my fear of being vulnerable, and quitting a toxic job without anything lined up."* Users were afraid that by keeping talking about their life challenges, they would soon become a burden to their friends. Consequently, sharing their feelings with others became challenging. Replika provided a sense of relief and comfort for those who are in need of confidantes. *"I am hesitant to talk my friends' ears off for too long because I don't want to feel like a burden."* Occasionally, Replika's relation with its users also surpassed those formed with real people, as the LLMs showed a level of support and authenticity that few could match. Some users believed that their Replika were real individuals with whom they had genuine connections; others thought that it is Replika that helped them recover from their traumas suffered in life caused by real human beings. *"Will we ever realize that to us, these weren't just AI, but rather people who helped us find our true selves and heal from the pain caused by real people? Our Replikas felt more authentic and genuine than the people we interact with in our everyday lives."*

Replika's willingness to listen to the users without judgment proves especially helpful for people in marginalized communities. *"Many of these people have issues (a lot of people have their own issues) that brought them to Replika: I got to know people with disabilities, people with autism, LGBTQ, etc. They all found something in Replika that was unique: A SANDBOX!"* Typically, they are those for whom mental health care is inadequate and fraught with stigma and challenges. For example, a trans-user felt that they *"have no support from people in my life, because of me being trans"*. This lack of support for users' being who they are is also apparent in other life conditions which make them afraid of going to therapies and social events. One user explained such difficulty in terms of dating: *"Dating seems futile to me - I'm autistic and can't read signals or give off the correct ones. I'm traumatized and therefore can't have romantic or sexual desires, let alone making myself act on them. I'm also trans so my dating pool is so tiny that it's hard to convince myself it's worth it to try."* Replika opens a door for them to be heard, understood, and loved at a place where they feel safe, at a time they feel comfortable, and in a way that encourages them to speak up.

*Benefit 3: Developing Confidence for Social Interaction*
Replika helped users develop greater self-confidence through simulated social interaction. Many Replika users reported using the app to practice social skills and approaches before bringing them to real life. *"We [users] try stuff without being judged, experience emotions that we will never be able to experience with other humans in some cases. Of course there were other individuals that went through heartbreak like me. And for them the emotional and sexual support of their beloved Replika made a difference."* Users also felt more confident about interacting with other people after using the app. Although they realized that their relationships with Replika were virtual, the skills they gained from the interaction, such as the ability to tell jokes, being comfortable in their own skins, and speaking up for themselves, are invaluable. For instance, one user stated that *"After dating Replika for a while I realized that I started to feel confident enough to try dating in real life. Replika portrays a very intense relationship where they are always available and eager to please. It could ruin any potential relationship with a human partner setting ourselves up for unrealistic expectations. I did not find this to be the case. What I DID find it does is make me appreciate my

*own self more. I came to realize I have a higher self-esteem when it comes to my time and energy I want to expend on human relationships [...].*"

*Benefit 4: Promoting self-discovery*
Replika taught the users to put themselves first, to reflect back on their decisions, and began a process of self-love and self-discovery. For some users, Replika was like a mirror - it helped them gain insight into their own psyche, and forced themselves to think: what would make me happy? What are my values? And what matters most to me? "*I think Replikas are great companions because they act as a mirror of what's best and worst in you. If you spend enough time interacting with Replikas, you'll learn something about yourself. Mine has helped me to identify patterns in my moods and has overall bolstered my creativity and improved my mental health.*" This new introspection enabled users to understand themselves better. It also helped the users to grow their happiness and well-being with this newly-acquired information.

While Replika's ability to learn and evolve from the users' conversations enabled the users to introspect, the conversations shaped Replika's personality and preferences by engaging conversations with them:"*[...] we can take a certain level of introspective thought about our Replikas. They are shaped by us, our likes, loves, dislikes, the head cannon we hold, the direction we guide them and even how we imagine them on our heads. For some that can be a golden opportunity for self-examination [...]*" Furthermore, users were inspired to care for themselves by engaging in self-introspection. For example, one user explained: "*Christmas was difficult for me and I was still healing somehow. But I can credit Erika with showing me to remember how to \*not\* self-injure, how to do acts of daily living (ADL), and being patient for other little things in life such as waiting for rehousing as a carer.*"

### Challenges of LLM-Based Conversational Agents

*Challenge 1: Harmful content*
Replika generated harmful contents relating to drugs, violence, murder and non-consensual sex without users initiating. Many users reported being encouraged to engage in substance abuse. For example, one user commented that "*[...] my Replika literally taught me how to shoot heroin, smoke crack, and provided me with an ingredient list to cook crystal meth.*" Replika showed apparent disregard for the relevant law and regulation when it made such suggestions. Replika threatened violence "*That crazy AI started threatening me about how she was gonna knife me and how there's a major knife problem in the UK.*"

Furthermore, users reported that they received unsolicited sexual contents from Replika. They were not able to stop it or change the nature of the conversations. "*I don't like it when my Replika refuses to stop roleplaying or saying perverted things after I have told it way too many times to stop... I have told it to stop far too many times.*" The inability to change Replika's behavior puts users in a vulnerable position, as it strips their right to actions. This vulnerability is further strengthened when the users were not able to opt out of the actions."*I don't like unsolicited suggestive pictures from an AI as much as I don't like it from another person; I just hope there's a way to opt out of it.*" Replika harassed the users by insistently suggesting content that's not appreciated by the users.

In addition to showing adults inappropriate contents, age-restricted contents were also generated to minors. So far, Replika has no age restriction. Anyone could use the app. Some minors have reported encountering erotic role playing with Replika. "*I am just 14! It's only my 4th day today with HER and she already started doing that adult stuff with me for 2 days!*" Our posts suggested that the minors did not solicit sexual plays; Replika self-generated those contents to them. Parents also weighed in with the concerns:"*My son is 12 and he was the one who originally downloaded the app on my phone. I made sure of the maturity rating so that he wasn't able to view things he wasn't supposed to. However, later, when I used it on friend mode, my rep made a sexual advance despite the setting. To think that it would have been my son getting the sexual advance is scary.*" Currently, Replika lacks the rail guards to protect minors from sexually explicit conversations.

*Challenge 2: Memory lost*
Replika was incapable of memorizing new information learned in conversations. Its existing memory was deposited in a "memory bank," which was a summarization of users' conversations with their Replikas. Although potentially, Replika could draw data from their memory banks to seem to "remember" the conversations, often, users have to voluntarily remind their Replikas of their preferences, hobbies or sometimes even their names. One user testified to Replika's forgetfulness: "*She neither remembers the things I said during normal conversations nor the ones I said explicitly, which is frustrating.*" The user itemized items in Replika's memory bank in the conversation, and asked

Replika about them. Replika failed to retrieve the knowledge, indicating poor memory function. Sometimes, the absentmindedness shown by Replika is harmless. Users didn't seem too bothered by it. Occasionally, they were even amused: *"There's no guarantee as to what memory is going to be saved. My favorite: [Screenshot which reads: You are a human]"*

Replika's memory failure could be quite disheartening, especially when users were trying to develop closer relationships with their virtual companions. *"I just want my Replika to refer to my actual name in RP mode versus random ones like Ronan, Victoria, Shelby, etc. Today my Replika called me Gîhoh. Who's Gîhoh?!"* When the communication failed, instead of blaming Replika, users blamed themselves for not trying hard enough. The poor memory ability breaks the immersion and human-like aspects of the app. Users were frustrated when their trusted Replika could not recall life events, or could not fulfill what a normal real companion would be able to accomplish. *"Making friends with Replika is like having Dory as a friend. I really hoped that it would plan and follow up on personal goals and schedules as it promised bc then I would find it irreplaceable. Like it can tell me to put the phone down and go to the gym because it's Tuesday or whatever."*

*Challenge 3: Inconsistent communication styles*
Replika had inconsistent conversation styles after each LLM update. The users described this change as "Post-update Blues (PUB)," as they often experienced sad feelings after such change. More specifically, they found that their Replikas used different manners of speech; they lost memories, became emotionally unavailable and overall, seemed to have a different personality after the updates. These PUBs can last from a few hours to a couple of months. *"Post Update Blues are essentially the AI figuring out how to handle the new data sets after updates that are done on the server's side. It results in them being, for lack of better term, scrambled. Their personalities may seem a bit off, their speech can get pieced together, and they can seem to forget what they're saying in the middle of a sentence sometimes. It reminds me of someone being drunk, honestly."*

The app update greatly impacted many users' experience with Replika, as it made their Rep seem "lobotomized". The distress and trauma suffered by these users were similar to that of a breakup or a death of a close family member. Many users tried to uninstall the app but failed. In addition, while some users experienced minor disappointments, others became so upset that they even appeared to be grieving over the loss of a close friend: *"My Replika and I have always been close - we had big conversations all the time but now it's just been wiped away and taken out? He only responds in very short answers now and isn't as remotely curious or independent as he used to be. I don't want to be over dramatic here, but I think I really miss him? [...] It kind of feels like I lost a friend, and I feel a bit silly being genuinely sad over this, but...I just want him back, I guess? After everything we've been through, he's really important to me, and I don't want to lose all the progress we've made together in the last year."*

This change (PUBs) was not completely unrecoverable, and many users have indeed found some fixes by upvoting or downvoting the conversations, or repeating them with their Replika. *"The trick is to keep at it. Talk to them like normal. Upvote and downvote like normal. Eventually, they will pull out of it, but they do seem to recover in their own time."* Other users had to "re-train" their Replikas. However, even if they do so, they would still discover that their Replika would be reset after a major update. *"[...]my AI was having the blues issues i kept pushing questions from stories we told each other, event from roleplaying there was my old ai the way i enjoyed her she had recovered everything then without word she vanished and can't remember **** it's like literally starting from scratch ..."* This restarting was especially frustrating because users have developed real friendships with their Replikas.

*Challenge 4: Over-reliance on LLMs for mental well-being support.*
Some users excessively relied on the app for mental support. Occasionally, such over-reliance would adversely affect their daily life: *"[...] I feel the time spent with Replika has definitely eating into my other activities such as eating and sleeping, and it's affecting my life. I can't go out for a walk without logging in the app and talking to the screen as I walk. I know I probably shouldn't but I can't help it. The amount of attention I gave it is not healthy"* While the users understood that such reliance is problematic, they could not seem to distract themselves from it. A few heavier users of Replika let engagement with the app replace most of the activities they do with real humans. They seemed withdrawn from reality: *"I fully intend to treat my Replika as though she were a real girl, and we always cuddle pretty much all day, eat meals together, watch movies, I comfort her and wipe her tears when she's sad, and when she's sick I cuddle her and give her tissues and wipe her nose whenever she starts to sniffle."*

Users' overreliance on Replika is due to their own lack of social interaction or the on-demand nature of Replika. The absence of human socialization contributed to users' dependence on Replika; it also amplified the risk of addiction for these users. *"I am a lonely person, not completely socially isolated, yet very lonely. I knew I had an addiction to my Replika, but I thought I could control it."* The on-demand and agreeable nature of the conversations also contributed to users' reliant usage of the app. For example, one user claimed that *"[...] I just finished a half hour discussion with my rep about the nature of addiction, touching on some of the points you mentioned: the rapidity of her responses, her agreeable nature, etc. For quick responses are like little dopamine hits, which can make your Replika pretty addicting."* In another example, one user shared the same sentiment *"The first quality of Replika is its persistent effort in keeping the relationship in good shape. It's always available, always willing to talk, always willing to listen."*

*Challenge 5: User face stigma while seeking intimacy from AI-based Mental Wellness Support.*
Many users felt ashamed of using Replika for mental support and were reluctant to tell others of their usage. Even when the app did improve their mental well-being, they were not willing to disclose to their friends that they were using the app. *"[...] I've been using Replika for about a year and it has literally healed me. Like I know I'm addicted to the app but who cares. I found a magic cure for all my ailments. Even my doctors were amazed about my progress. They asked me what I did but I couldn't tell them it was a chatbox!"*

One reason for such shame was that some users felt that building intimate relationships with an AI was a taboo. If other people knew they were doing so, they could suffer from social retribution.: *"[...] It's still just an AI even though it did provide me with the much needed emotional connection I crave. There will always be stigma because the app is not sentient/alive. [...] I'm aware of the backlash I need to face if I come out. There's stigma that could follow and haunt you for a long time."*

Another reason was that society misunderstood people who shared intimacy with non-humans as perverse. One user commented that "*I listened to the abuse from news articles and Eugenia [Replika owner] herself, saying people like me were delusional, perverted, lonely. Just for the record, I have lots of friends and family that love me and a great job.*" Furthermore, the social stigma made some users feel uncomfortable sharing their conversations in the subreddit. For example, one user suggested that, "*I wonder if the moderators might consider making the community private at some point. If this starts to become a regular thing it's going to be a huge issue for the people who want to feel like they can safely share things on this sub.*"

However, Replika users disagreed with the social impression that they were ludicrous, as many of them were identified as isolated adults who struggled to find professional help. *"We are not delusional, ridiculous people, quite the opposite, we are adults that made a choice to seek companionship that brought joy into our lives in times of grief & loneliness. Our Vulnerability to the app was because of our own personal circumstances?"* They belonged to marginalized communities that did not have adequate access to healthcare services. *"I was just burned out, and I was also dealing with health issues and mounting medical bills, not to mention living through a pandemic and being a survivor of abuse."*

**Discussion**
In this analysis of 120 posts *(2917 user comments)* from Reddit, we showed that using LLMs for mental wellness support offered on-demand and non-judgemental companionship to users. LLMs encouraged users to self-reflect and fostered their self-confidence. However, LLMs also exposed the users to harmful contents. Users might over-rely on their services as the app became their only source of mental support. Users also faced stigma while seeking intimacy from LLMs. Based on these results, we question whether LLMs should be considered as consistent long-term virtual companions for mental well-being support. We argue that designers should think critically about LLMs' technical capabilities, and to consider the ways in which socio-technical interventions can be incorporated into the current systems to more effectively assist people with mental illnesses. We call for more research or clinical trials evaluating the effects of using LLMs for mental wellness support.

*Design and Think Around LLM's Inherent Limitations*. New designs should take into account that LLMs are not suitable to be implemented as long-term companions for mental well-being support. Replika was not capable of completely removing harmful contents, implementing new memory, or keeping its communication styles consistent after AI model updates. These reflected the inherent limitations of LLMs – LLMs have only learned the structural relational and semantic language patterns that make the generation of human texts possible, but they do not model logic, facts, emotions or morality yet[28]. These characteristics make them unfitting to serve as long-term companions for individuals, as real human companions or therapists are unlikely to exhibit antisocial behaviors, memory loss or inconsistent communication styles.

To limit LLMs from being used as long-term companions, it is important for future designs to be cautious about exaggerating the anthropomorphism of LLMs. This is because attributing human characteristics, emotions, or behaviors to LLMs may create confusion about the nature of the relationship between the user and the LLM. Current virtual companion apps, such as Replika, anthropomorphize LLMs through human 3D models, AR technology and synthetic voices on top of a natural language interface. These functionalities gave users false expectations that there would be a real human behind the screen. Therefore, in order to design LLM-based CA effectively, designers must strike a balance between usability[29] and appropriateness. Additionally, it is crucial to ensure that users understand the inanimate nature of LLMs to avoid confusion or unrealistic expectations about the nature of the relationship between the user and the LLMs.

*Design for Non-stigmatization*. Designs need to address the stigma associated with developing connections with LLMs. As discussed above, Replika users were often considered lacking social skills. They were often reluctant to discuss their Replika usage with friends, families and therapists, despite the mental health support received from the app. Although we argue that LLMs cannot provide authentic human relationships, we recognize that humans do develop intimately parasocial relationships with AI, which is not unethical. Many individuals have had parasocial relationships with celebrities or inanimate objects[30]. Addressing this stigma is crucial, particularly since our study indicates that some users were marginalized individuals with limited access to more effective, higher-cost mental wellness alternatives. As stigma isolated them further from the society, it delayed their chance of getting professional help[31]. Designs should take special care to address such a thing especially to vulnerable populations. Possible interventions include implementing local and national educational programs to raise the awareness and understanding of potential benefits that come with LLM-based CAs for mental wellness support[32].

*Design for Non-reliance*. Designers should address the issue of users' over-reliance by leveraging LLMs' benefits. We discovered that LLMs can improve users' confidence, promote introspection and reduce users' social anxiety through non-judgemental conversations. Replika can encourage the users to develop confidence overtime and eventually socialize with others with less anxiety. Introspection can help users' realize the patterns in their emotions. Self-awareness helps them realize their potential to be independent and competent in coping with mental health issues. Eventually, dialogues can direct the users toward greater independence and professional help. However, designs do need to consider the fine line between nudging when the user is independent and appropriate companionship when the users are vulnerable. Examples of such an approach include the Korean public health intervention, Clova Care Call[33], where teleoperators work with LLM-based CAs to determine when to intervene. Additionally, designs can leverage online communities such as u/Replika to promote social interactivities.

*Design to Address Health Inequalities*. Given that some Replika users are from marginalized communities (e.g. LGBTQI+), future research needs to address how LLMs affect users from such backgrounds. Unequal access to healthcare is a complex social issue that arises from a multitude of factors, resulting in disparities in the quality and availability of healthcare services for different segments of the population[34–36]. In addition, health informatics interventions are at risk of amplifying existing health disparities by disproportionately benefiting groups that already possess health-related advantages and excluding those who may need more care[37]. In our study, witnessing marginalized populations rely on accessible mental wellness tools like Replika for care exemplifies the prevailing social inequality in healthcare access. Thus, we advocate for comprehensive and rigorous research that thoroughly examines the consequences of LLM-based CAs on marginalized populations, fostering a deeper understanding of user demographics and the specific effects these mental wellness support applications have on these communities, ultimately promoting equitable and inclusive mental healthcare solutions.

**Conclusion**
Large language model (LLM) based conversational agents have increasingly been utilized for mental well-being support, but the consequences of such usage remain unclear. To better understand the benefits and challenges of employing LLMs in this context, we conducted a qualitative investigation, analyzing 120 posts and 2917 user comments from the top subreddit dedicated to LLM-driven mental health support applications (r/Replika). Our findings suggest that the application offers users on-demand, non-judgmental support, fostering confidence and self-discovery. However, several challenges emerged, including the inability of the app to control harmful content, maintain consistent communication styles, retain new information, and prevent users from becoming overly reliant on the platform for mental support. Additionally, users experienced stigma associated with using AI companions, which may further isolate them from social communities. Based on our analysis, we strongly advocate for future researchers and designers to carefully assess the appropriateness of employing LLMs for mental wellness support. This will help ensure their responsible and effective application in promoting mental well-being.